\documentstyle[11pt]{article}

\newcommand {\be}{\begin{equation}} 
\newcommand{\fe}{\end{equation}}
\newcommand{\eqn}{\label}
\newcommand{\bel}{\begin{equation}\label}
\def\thf{\baselineskip=\normalbaselineskip\multiply\baselineskip
by 7\divide\baselineskip by 6}            \thf

\def\spose#1{\hbox to 0pt{#1\hss}}\def\lta{\mathrel{\spose{\lower 3pt\hbox
{$\mathchar"218$}}\raise 2.0pt\hbox{$\mathchar"13C$}}}  \def\gta{\mathrel
{\spose{\lower 3pt\hbox{$\mathchar"218$}}\raise 2.0pt\hbox{$\mathchar"13E$}}}

\begin{document}

\title{HAS THE BLACK HOLE EQUILIBRIUM PROBLEM BEEN SOLVED?}

\author{B. CARTER}

\date{D.A.R.C., (UPR 176, CNRS),
\\ Observatoire de Paris, 92 Meudon, France }

\maketitle

{\bf Abstract. \ } When the term ``black hole'' was originally coined
in 1968, it was immediately conjectured that the only pure vacuum
equilibrium states were those of the Kerr family. Efforts to confirm
this made rapid progress during the ``classical phase'' from 1968 to
1975, and some gaps in the argument have been closed during more recent
years. However the presently available demonstration is still subject
to undesirably restrictive assumptions such as non-degeneracy of the
horizon,  as well as analyticity and causality in the exterior.

\vskip 1 cm

\section{Introduction}
\label{Section1}

The purpose of this report is to present a brief overview of the
present state of progress on the still not completely solved problem of
the classification of black hole equilibrium states within the
(astrophysically motivated) context of Einstein's pure vacuum theory and
its electrovac generalisation in ordinary four dimensional spacetime.

In recent years there has been a considerable resurgence of
mathematical work on black hole equilibrium states, but most of it has
been concerned with more or less exotic generalisations, not restricted to
four dimensions,  involving speculative extensions of
Einstein's theory to allow for inclusion of various scalar and other
(e.g. Yang Mills type) fields such as those ocurring in low energy
limits of superstring theory. The present review will not even attempt
to deal with  this rapidly developing and open ended area of
investigation. The not quite so fashionable but -- as far as 
the observable physical world is concerned -- more
soundly motivated subject of black hole equilibrium states with
surrounding rings of matter (such as would result from accretion from
external sources) is also beyond the scope this article.

One reason why there has not been so much recent work on what, from an
astrophysical point of view, is the most important problem in black hole
equilibrium theory, namely that of the pure vacuum states in four dimensions,
is the widespread belief  that the problem was solved long ago, and that the
solution is just what was predicted by my original 1967
conjecture~\cite{Carter67,Carter68}, i.e. that it is completely provided by
the subset of Kerr solutions~\cite{Kerr63} for which $a^2\leq M^2$ where
$a=J/M$  is the ratio of the angular momentum $J$ to the mass $M$. This belief
rapidly gained general acceptance in astrophysical circles when -- following
the example of Israel's earlier 1967 work~\cite{Israel67} providing strong
evidence that (as has since been confirmed) the only strictly static (not just
stationary) solutions were given by the special Schwarzschild ($J=0$) case --
I was able~\cite{Carter71} in 1971 to obtain a line of argument that provided
rather overwhelming, though by no means absolutely watertight, mathematical
evidence to the effect that the most general solution is indeed included in
the Kerr family.

Many of the interested parties, particularly observationally motivated
astrophysicists, considered that the conversion of the original
plausibility argument into an utterly unassailable mathematical proof
was merely a physically insignificant technical formality, that could
be left as an exercise for the amusement of obsessively rigoristic pure
mathematicians.  However such lack of interest was not the only reason
why subsequent progress on the problem has been rather slow. It  was
soon shown by those -- starting with Hawking~\cite{Hawking72,Hawking73}
-- who took the question seriously after all that the mathematical work
needed to deal with the various apparently small technical gaps and
loose ends in the comparitively simple 1971
argument~\cite{Carter71,Carter73} is harder than might have been hoped.
Thus despite the very considerable efforts of many people -- of whom
some of the most notable after
Hawking~\cite{Hawking72,Hawking73,HawkingEllis73} have been 
Robinson~\cite{Robinson74,Robinson75,Robinson77},
followed (for the electrovac generalisation) by
Bunting~\cite{Bunting83,Carter85} and Mazur~\cite{Mazur82,Mazur84}), and
most recently (since my last comprehensive review of the
subject~\cite{Carter87}) Wald and his collaborators
~\cite{SudarskyWald91,SudarskyWald93,ChruscielWald93,ChruscielWald94} --
there still remains a lot that needs to be done before we shall have
what could be considered a mathematically definitive solution even of
the pure vacuum problem, not to mention the more formidable challenge
of its electrovac generalisation.

It will be convenient to present the results in 
chronological order, which roughly corresponds to that of their
logical development except for a few cases in which newer work has
provided more elegant methods of rederiving results that were orginally
obtained by more laborious means. Section 2 rapidly recalls some of
the relevant results (culminating in Israel's theorem) obtained during the
prolonged period of general confusion that I refer to as the ``preclassical
phase'', prior to the introduction of the term ``black hole'' and to the
general recognition that the disciples of  Ginzburg and Zel'dovich had been
right~\cite{Thorne94} in arguing that what it represents is a generic
phenomenon -- not just an unstable artefact of spherical symmetry as many
people, including Israel himself~\cite{Israel87}, had speculated. Section 3
describes the rapid progress made during what I refer to as the ``classical
phase'' -- the beginning of what Israel~\cite{Israel87} has referred to as the
``age of enlightenment''-- immediately following the definitive formulation of
the concept of a ``black hole'' (in terms of the ``outer past event horizon''
in an asymptotically flat background) so that the corresponding equilibrium
state problem could at last be posed in a mathematically well defined form.
Section 4 describes the substantial though slower progress that has been made
in what I refer to as the ``postclassical phase'', that began  when
the main stream of work on black holes had been diverted to quantum aspects
following the discovery of the Hawking effect~\cite{Hawking75}.

The final section  draws the intention of newcomers to the field, for
whom this article is primarily intended, to the mathematically
challenging (even if physically less important) problems that still
remain to be tackled: these include not only the questions concerning
the technicaly awkward degenerate limit case and the assumptions about
spherical topology and analyticity that have been discussed by
Chrusciel~\cite{Chrusciel94} and also, in a very extensive and up to
date review, by Heusler~\cite{Heusler96}, but also the largely neglected
question of the assumption of causality, i.e. the absence of closed
timelike curves.

A propos of the latter, is ironic that while providing a fascinating
account of the mental blockages that impeded earlier workers in the
theory of both black holes and time machines (i.e.  regions of
spacetime threaded by closed timelike curves) Thorne's recent history
of ``Black holes and time warps''~\cite{Thorne94} has a blind spot of
its own, in that it discusses only the kind of closed timelike curve
whose presence depends on topological multiconnectedness in
``wormholes'' of a rather artificial kind (so that the resulting
causality violation is ``trivial''~\cite{Carter78} in the sense of being
in principle removable by replacing the spacetime model by its locally
equivalent universal covering space). What is rather surprising is
Thorne's failure to mention the kind of time machine
exemplified~\cite{Carter68} by the Kerr solutions for $a^2>M^2$, in
which causality violation of a more ``flagrant''~\cite{Carter78} (not so
easily removable) kind occurs.  In the Kerr black hole case $a^2\leq
M^2$ the causality violation is confined to the interior, but the
unsolved problem is whether there exist other black hole equilibrium
solutions in which such causality violation occurs {\it outside} the
horizon. The formal existence of such pathological black hole solutions
might of course be reasonably supposed to be irrelevant for realistic
physical purposes.  However the same kind of objection could be raised
to Thorne's ``wormhole'' time machines: if the latter are nevertheless
at least of sufficient mathematical interest to be worth investigating
then the same applies {\it a fortiori} to black hole time machines if
they exist, a possibility that is by no means excluded by any of the
work carried out so far.

\section{The preclassical phase (1915-67).}
\label{Section2}

What I refer to as the preclassical phase in the development of black
hole theory is the period of unsystematic accumulation of more or less
relevant results prior to the actual use of the term ``black hole''.
 This period began with the discovery in 1916 by
Schwarzschild~\cite{Schwarz16} of his famous asymptotically flat vacuum
solution, whose outer region is strictly static, with a hypersurface
orthogonal timelike Killing vector whose striking feature is that its
magnitude tends to zero on what was at first interpreted as a spacetime
singularity, but was later recognised to be interpretable as a regular
boundary admitting a smooth extension to an inner region where the
Killing vector becomes spacelike.  This preclassical phase culminated
in Israel's 1967 discovery~\cite{Israel67} of a mathematical argument to
the effect that the Schwarzschild solution is uniquely characterised by
these particular properties, i.e.  the original spherical example is
the only example.  The significance of this discovery was the subject
of an intense debate that precipitated the transition to  the
``classical phase'', inaugurating what Israel~\cite{Israel87} has termed
the ``age of enlightenment'', which dawned when the preceeding
confusion at last gave way to a clear concensus.  Thrilling eyewittness
accounts of the turbulent evolution of ideas in the ``golden age'' of
rapid progress, during the transition from the preclassical to the
classical phase, have been given by Israel himself~\cite{Israel87} and
from a different point of view by Thorne~\cite{Thorne94}, while a
historical account of the more dilatory fumbling in the early years of
the preclassical phase has been given by
Eisenstaedt~\cite{Eisenstaedt82}.

During most of the ``unenlightenned'' preclassical period, from 1915
until about 1960, nearly all the relevant work, starting with that of
Schwarzschild, was in fact based on the simplifying postulate of
spherical symmetry.  An important consequence of this restriction was
demonstrated by Birkhoff's 1923 theorem~\cite{Birkhoff23}, which showed
that the staticity property used in Schwarzschild's derivation of his
solution need not have been postulated indendently of the spherical
symmetry, since it followed as an automatic consequence of the vacuum
field equations.  An important step towards the concept of what
would be called a ``black hole'' was the analysis~\cite{Oppy39} of the
gravitational collapse of pressure free matter by Oppenheimer and
Snyder in 1939. However in the pure vacuum case, on which the present
review is focussed, progress stayed remarkably slow for a very long
time, and people remained confused by the special limit in the
Schwarzschild solution where the circumferencial radius $r$ reaches the
value $2M$ (in relativistic units). Despite the construction of
analytic extensions beyond this limit by many earlier
workers~\cite{Painleve21,Gullstrand22,Eddington24,Lemaitre30,Synge50,Finkelstein58}
a clear understanding was obtained only after more complete extensions
were made by Frondsal~\cite{Frondsal59}, Kruzkal~\cite{Kruzkal60}, and
Szekeres~\cite{Szekeres60}.

Due to a renaissance of interest  (following the observational
discovery of quasars) progress was much more rapid during what
Thorne~\cite{Thorne94} has referred to as the ``golden age'', which
began during the last half dozen years of the preclassical phase and
continued through what I call the classical phase (ending when most of
the easiest problems had been solved at about the time Hawking diverted
attention attention to less astrophysically relevant quantum effects).
It was during the late ``golden'' period of the pre-classical phase,
roughly from 1961 to 1967, that results of importance for vacuum black
hole theory began to be obtained without reliance on a presupposition
of spherical symmetry.  The most important of these results were of
course Kerr's 1963 discovery~\cite{Kerr63} of the family of stationary
asymptotically flat vacuum solutions characterised by a degenerate
``type D'' Weyl tensor, and the 1967 Israel theorem~\cite{Israel67}
referred to above.

As well as these two specific discoveries, the most significant
development during this final ``golden'' period of the preclassical
phase was the animated debate -- under the leadership of Ginzburg and
Zel'dovich ~\cite{ZeldovichNovikov67} in what was then the Soviet Union,
of Wheeler and later on Thorne~\cite{HTWW65} in the United States, and
of Sciama and Penrose~\cite{Penrose65} in Britain -- from which the
definitive conceptual machinery and technical jargon of black hole
theory finally emerged.  Prior to the discovery of the Kerr
solution~\cite{Kerr63}, when the only example considered was that of
Schwarzschild, it had not been thought necessary to distinguish what
Wheeler later termed the ``ergosphere'' -- where the Killing vector
generating the stationary symmetry of the exterior ceases to be
timelike -- from the ``outer past event horizon'' bounding what Wheeler
later termed the ``black hole'' region, from which no future timelike
trajectory can escape to the asymptotically flat exterior.  In its
original version, Israel's 1967 theorem~\cite{Israel67} (as well as its
electrovac generalisation~\cite{Israel68}) was effectively formulated in
terms of an ``infinite redshift surface'' that was effectively taken to
be what in strict terminology was really the ``ergosphere'' rather than
the ``outer past event horizon'': this meant that the significance of
the theorem  for the theory for what were  to be called a ``black
holes'' was not clear until it was understood that (as shown in my
thesis~\cite{Carter67,Carter78} and pointed out independently by
Vishveshwara~\cite{Vishu68})  subject to the condition of strict
staticity postulated by Israel (but not in the more general stationary
case exemplified by the non-spherical Kerr~\cite{Kerr63} solutions) the
the ``outer past event horizon'' actually will coincide with the
ergosphere.

One of the first to appreciate the distinction between (what would come to be
known as) the horizon and the ergosphere, and to recognise the members of the
relevant ($a^2\leq M^2$) Kerr subset as prototypes of what would come to be
known as black hole solutions was Boyer~\cite{Boyer65,Boyer67,Boyer69}.
However at the time of the first detailed geometrical investigations of the
Kerr solutions~\cite{Carter66,Boyer67} (the purpose for which, following a
suggestion by Penrose, I originally introduced the scheme of representation by
the kind of  conformal projection now commonly known as a ``Penrose
diagram''), it was assumed by Boyer and the others involved, including myself,
that we were dealing just with a particularly simple case within what might
turn out to be a much more extensive category. However the publication of the
1967 Israel theorem~\cite{Israel67} -- which went much of the way towards
proving that the Schwarzschild~\cite{Schwarz16} solution is the only strictly
static example -- immediately lead to the question of whether the Kerr
solutions might not be similarly unique.

The explicit formulation of this suggestion came later to be loosely
referred to in the singular as the ``Israel-Carter conjecture'', but
there were originally not one but two distinct versions. The stronger
version -- suggested by the manner in which the Israel theorem was
originally formulated -- conjectured that the relevant Kerr subfamily
might be the only stationary solutions that are well behaved outside
and on a regular ``infinite redshift surface'' -- a potentially
ambiguous term that in the context of the original
version~\cite{Israel67} of Israel's theorem effectively meant what was
later to be termed an ``ergosurface'' rather than an ``event
horizon''.  The weaker version, first written unambiguously in my 1967
thesis~\cite{Carter67,Carter68}, conjectured that the relevant Kerr
subfamily might be the only stationary solutions that are well behaved
all the way in to a regular black hole horizon, not just outside the
ergosphere.

Work by Bardeen~\cite{Bardeen73} and others on the effects of stationary
orbitting matter rings (which can occur outside the horizon but inside
the ergosphere of an approximately Kerr background) soon made it
evident that strong version of the conjecture is definitely wrong, no
matter how liberally one interprets the rather vague qualifications
``regular'' and  ``well behaved''. On the other hand the upshot of the
work to be described in the following sections is to confirm the
validity of my weaker version, as expressed in terms of the horizon
rather than the ergosphere. It is nevertheless to be remarked that, as
was pointed out by Hartle and Hawking~\cite{HartleHawking72}, the
generalisation of this conjecture from the Kerr pure vacuum
solutions~\cite{Kerr63} to the Kerr-Newman electrovac solutions~\cite{Newman65}
is not valid, since the solutions due to Papapetrou~\cite{Papa45} and
Majumdar~\cite{Majumdar47} provide counterexamples.  It is also to be
emphasised that, as will be discussed in the final section, the question still
remains entirely open, even in the pure vacuum case, if the interpretation of
the qualification ``well behaved'' is relaxed so as to permit causality
violation outside the horizon of the kind that is actually
observed~\cite{Carter68} to occur in the inner regions of the Kerr examples.

\section{The classical phase (1968-75).}
\label{Section3}

What I refer to as the classical phase in the development of black hole
theory began when the appropriate  conceptual framework and the
corresponding generally accepted technical terminology became
available, facilitating clear formulation of the relevant mathematical
problems, whose solutions could then be sought by systematic research
programs, not just by haphazard approach of the preclassical period.
The relevant notions had already began to become clear to a small
number of specialists (notably Wheeler's associates, including Thorne
and Misner, in the United States, and Penrose's associates, including
Hawking and myself in Britain) during the period of accelerated
activity at the end of what I call the preclassical phase.

However it was not just theoretical progress that precipitated the
rather sudden (``first order'') transition to what I call the classical
phase.  Just as it was the discovery of the quasar phenomenon that
stimulated the ``golden age''~\cite{Thorne94} of rapid progress, so
also, rather similarly, it was another observational event, namely the
accidental discovery of pulsars by Bell and Hewish, that inaugurated
the transition from the ``preclassical phase'' to the ``age of
enlightenment''~\cite{Israel87} during the second half of the ``golden
age''.  Unlike the quasar phenomenon, whose underlying mechanism is far
from clear even to day, the pulsar phenomen was rapidly elucidated: in
the early months of 1968 it was already generally generally recognised
to be attributable to neutron stars, whose likely existence had long
been predicted by theoreticians, but whose reality had never until then
been taken very seriously by the majority of the astronomical
community.  The 1968 confirmation that neutron stars definitely exist
and are directly observable immediately transformed the status of
theoreticiens in the eyes of the observers. (Prior to 1968 even our
firmest affirmations were treated with the greatest scepticism; after
1968 even our most tentative speculations, as well as our conjectures
about ``black holes'', were received as oracular pronouncements.)
This meant that the beginning of the ``classical phase'' was characterised
not only by the establishment of an ``enlightenned'' consensus among the
previously disparate groups of specialists working in the field, but also by
the recognition for the first time by a much wider public that a new and
important field of theoretical astrophysics had been born. At the beginning
of 1968 the term ``black hole'' was known only to a handful of participants
in the seminars organised at Princeton by Wheeler; by the end of 1968 the
term had already been widely publicised in televised science fiction so that
it was already known (if not understood) by millions of people all over the
world.

At a time when the existence of black holes produced by burnt out stars
throughout our galaxy and others was already widely albeit prematurely
recognised by much of the astronomical community,  it became urgent for
the theoreticians actually working in the field to settle the question
of the physical relevance of the black hole scenario, which requires
that it should occur not just as an unstable special space (which was
the implication that Israel was at first inclined to draw from his
theorem~\cite{Israel87}) but as a generic phenomenon {as Zel'dovich and
his collaborators had been claiming~\cite{Thorne94}).

The strongest conceivable confirmation of the general validity of the
black hole scenario is what would hold if Penrose's 1969 cosmic
censorship conjecture~\cite{Penrose69} were valid in some form.
According to this vaguely worded conjecture, in the framework of a
``realistic'' theory of matter the singularities resulting (according
to Penrose's earlier ``preclassical'' closed trapped surface
theorem~\cite{Penrose65}) from gravitational collapse should generally
be hidden within the horizon of a black hole with a regular exterior.
However far from providing a satisfactory general proof of this
conjecture, subsequent work on the question (of which there has not been 
as much as would be warranted) has tended to show that can be valid only if
interpreted in a rather restricted manner. Nevertheless, despite the
construction of various more or less artificial counterexamples by
Eardley, Smarr, Christodoulou and others~\cite{Eardley87} to at least the
broader interpretations of this conjecture, it seems clear
that there will remain an extensive range of ``realistic''
circumstances under which the formation of a regular black hole
configuration is after all to be expected.

It remains a controversial question (and in any case one that is beyond
the scope of this discussion of pure vacuum equilibrium states) just
how broad a range of circumstances can lead to regular black hole
formation, and whether or not ``naked singularities'' can sometimes be
formed instead under ``realistic'' conditions.  However that may be,
all that is actually needed to establish the relevance of black hole
for practical physical purposes (as a crucial test of Einstein's
theory, and assuming the result is positive, as an indispensible branch
of astrophysical theory) is the demonstration of effective stability
with respect to small purturbations of at lease some example. This
essential step was first achieved in a mathematically satisfactory
manner for the special case of the original prototype black hole
solution, namely the Schwarzschild solution, in a crucially important
quasi-normal mode analysis~\cite{Vishu70} by Vishveshwara in 1970.
Another important article~\cite{Price72} by Price provided a more
detailed account of the rate at which the solution could be expected to
tend towards the Schwrzschild form under realistic circumstances as
seen from the point of view of an external observer. The work of
Vishveshwara and Price put the physical relevance of the subject beyond
reasonable doubt by demonstrating that this particular (spherical)
example is not just stable in principle but that it will also be
stable in the practical sense of tending to its stationary (in this
particular case actually static) limit within a timescale that is
reasonably short compared with other relevant processes: in the
Schwarzschild case the relevant timescale for convergence, at a given
order of magnitude of the radial distance from the hole, turns out just
to be comparable with the corresponding light crossing timescale.

During the remainder of the ``classical phase'', important 
work~\cite{Teukolsky72,Teukolsky73,PressTeukolsky73})
by Teukolsky, Press and others made substantial progress towards the 
confirmation that the Kerr solutions are {\it all} similarly stable 
so long as the specific angular momentum parameter $a=J/M$ is
less than its maximum value, $a=M$. However the possibility that instability
might set in at some intermediate value in the range $0<a<M$ was not
conclusively eliminated until much later on, in the ``post-classical'' era,
when (following a deeper study of the problem by Kay and
Wald~\cite{KayWald87}) the question was settled more conclusively
by the publication of a powerful new method of analysis developed by
Whiting~\cite{Whiting89}.

While this work on the stability question was going on, one of the main
activities characterising the ``classical phase'' of the subject was
the systematic investigation (along lines
pionneered~\cite{Christodoulou70,ChristodoulouRuffini71} by
Christodoulou and Ruffini) of the general mechanical laws governing the
behaviour of stationary and almost stationary black hole states. Work
by a number of people including Hawking, Hartle, Bardeen, and
myself~\cite{Hawking72,Carter72,HawkingHartle72,Hartle73,BCH73} (and later, as
far as the electromagnetic
aspects~\cite{Carter73,Znajek78,Damour78,Carter79} are concerned, also
Znajek and Damour) revealed a strong analogy with the thermodynamical
behaviour of a viscous (and electrically resistive) fluid. (Following a
boldly imaginative suggestion by Bekenstein~\cite{Bekenstein73}, the
suspicion that this analogy could be interpreted in terms of a deeper
statistical mechanical reality was spectacularly
confirmed~\cite{Hawking75} when Hawking laid the foundations of quantum
black hole theory.)

It was the substantial theoretical framework built up in the way during the
``classical'' phase that decisively confirmed the crucial importance of the
equilibrium state problem on which the present article is focussed.
Returning to this more specialised topic, I would start by recalling that
-- as well as the provision of the first convincing demonstration that
(contrary to what Israel~\cite{Israel87} had at first been incined to
suspect) a black hole equilibrium state can indeed be stable -- a noteworthy
byproduct of Vishveshwara's epoch making paper~\cite{Vishu70} was its
analysis of stationary as well as dynamical perturbations, which provided
evidence favorable to my  uniqueness conjecture~\cite{Carter67,Carter68} in
the form of a restricted ``no hair'' theorem to the effect that the only
stationary pure vacuum generalisations obtainable from a Schwarzschild black
hole by infinitesimal parameter variations are those of the Kerr family (in
which relevant small parameter is the angular momentum $J$).

Encouraged by Vishveshwara's confirmation~\cite{Vishu70} of the
importance of the problem, I immediately undertook the first systematic
attempt~\cite{Carter71,Carter73} at verification of my uniqueness
conjecture for the $a^2\leq M^2$ Kerr solutions as stationary black
hole states. For the sake of mathematical simplicity I restricted my
attention to the case characterised by spherical topology and axial
symmetry, conditions that could plausibly be guessed to be
mathematically necessary in any case. I also ruled out consideration of
conceivable cases in which closed timelike or null lines occur outside
the black hole horizon (not just inside as in in the Kerr
case~\cite{Carter68}), a condition that is evidently natural on
physical grounds, but that is not at all obviously justifiable as a
mathematically necessity.

Within this framework I was able in 1971 to make two decisive steps forward,
at least for the generic case for which there is a non zero value of the
decay parameter $\kappa$ which (in accordance with the ``zeroth'' law of
black hole thermodynamics~\cite{BCH73}) must always be constant over the
horizon in a stationary state. The first of these steps~\cite{Carter71} was
the reduction of the four dimensional vacuum black hole equilibrium problem
to a two dimensional non-linear elliptic boundary problem, for which the
relevant boundary conditions involve just two free parameters: the outer
boundary conditions depend just on the mass $M$ and the inner boundary
conditions depend just on the horizon scale parameter, $c$ which is
proportional to the product of the decay parameter $\kappa$ with the horizon
area ${\cal A}$. The precise specification of this parameter $c$ (originally
denoted by the letter $b$, and commonly denoted in more recent litterature
by the alternative letter $\mu$) is given by the definition $c=\kappa{\cal
A}/4\pi$, and its value in the particular case of the Kerr black holes is
given the formula $c=\sqrt{M^2-a^2}$ with $a=J/M$.

The second decisive step obtained in 1971 was the
demonstration~\cite{Carter71} that the two dimensional boundary problem
provided by the first step is subject to a ``no hair'' (i.e. no
bifurcation) theorem to the effect that within a continuously
differentiable family of solutions (such as the Kerr family) variation
between neighbouring members is fully determined just by the
corresponding variation of the pair of boundary value parameters, i.e.
the solutions belong to disjoint 2-parameter families in which the
individual members are fully specified just by the relevant values of
$M$ and $c$. The only known example of such a family was the Kerr
solution, which of course includes the only spherical limit case,
namely that of Schwarzschild. The theorem therefore implied that if,
contrary to my conjecture, some other non-Kerr family of solutions
existed after all, then it would have the strange property of being
unable to be continuously varied to a non-rotating spherical limit. On
the basis of experience with other equilibrium problems this strongly
suggested that, even if other families did exist, they would be
unstable and therefore physically irrelevant, unlike the Kerr solutions
which, by Visveshwara's work~\cite{Vishu70} were already known to be
stable at least in the neighbourhood of the non-rotating limit.

Having drawn the conclusion from this plausible but debateable line of
argument that for practical astrophysical purposes a pure vacuum black
hole equilibrium could indeed be safely presumed to be described by a
Kerr solution, I turned my attention to the problem of generalising
this argument from the pure vacuum to the electrovac case. In the
degenerate ($\kappa=0$) case, it had been pointed out by Hartle and
Hawking~\cite{HartleHawking72} that the Kerr-Newman~\cite{Newman65}
family did not provide the most general equilibrium solution, due to
the existence of counterexamples provided by the Papapetrou-Majumdar
solutions, but it remains plausible to conjecture that the most general
non-degenerate solutions are indeed provided by the Kerr-Newman family
(whose simple spherical limit is the Reissner-Nordstrom solution).  The
electrovac generalisation of the first step of my 1971
argument~\cite{Carter71} turned out to be obtainable without much
difficulty~\cite{Carter73}, the only difference being that the ensuing
two-dimensional non-linear boundary problem now involved two extra
parameters representing electric charge and magnetic monopole. As in
the pure vacuum case, an essential trick was the use of a modified
Ernst~\cite{Ernst68} transformation based on the {\it axial} Killing
vector (instead of the usual time translation generator) so as to
obtain  a variational formulation for which -- assuming {\it causality}
-- the action would be {\it positive definite}.  The second step was
more difficult: I did not succeed in constructing a suitable electrovac
generalisation of the divergence identity -- with a rather complicated
but (like the action) {\it positive definite} right hand side -- that
had enabled me to establish the pure vacuum ``no hair''
theorem~\cite{Carter71} for axisymmetric black holes in 1971, but an
electrovac identity of the required -- though even more complicated --
form was finally obtained by Robinson~\cite{Robinson74} in 1974.

While this work on the electromagnetic generalisation was going on, a deeper
investigation of the underlying assumptions was initiated by
Hawking~\cite{Hawking72,Hawking73,HawkingEllis73}, who made very important
progress towards confirmation of the supposition that the topology would
be spherical, and that the geometry would be axisymmetric. The latter was
achieved by I call the ``strong rigidity'' theorem, which was originally
advertised~\cite{HawkingEllis73} as a demonstration that -- assuming
analyticity -- the black hole equilibrium states would indeed
have to be axisymmetric (and hence by my earlier ``weak rigidity''
theorem~\cite{Carter72} {\it uniformly} rotating) except in the static case,
which in the absence of external matter was already known -- from the recent
completion~\cite{MRS73,MRS74} of the program initiated by
Israel~\cite{Israel67,Israel68} -- to be not just axisymmetric but geometrically
(not just topologically) spherical.

The claim to have adequately confirmed the property of
axisymmetry~\cite{HawkingEllis73} was however one of several
exagerations and overstatements that were too hastily put forward
during that exciting ``classical'' period of breathlessly rapid
progress. In reality, all that was mathematically established by the
``strong rigidity'' theorem was just that in the non axisymmetric case
the equilibrium state would have to be ``non-rotating'' (in the
technical sense that is explained in the appendix).  The argument to
the effect that this implied staticity depended on Hawking's
generalisation~\cite{Hawking72} of the original
Lichnerowicz~\cite{Lichnerowicz55} staticity theorem, which in turn
assumed the absence of an ``ergosphere'' outside the (stationary
non-rotating) horizon -- a litigious supposition whose purported
justification in the non-rotating case was based on heuristic
considerations~\cite{HawkingEllis73} that have since been recognised to
be fundamentally misleading, due to the existence of counterexamples. A
satisfactory demonstration that the non-rotating case must after all be
static was not obtained until the comparitively recent
development~\cite{SudarskyWald91,ChruscielWald93,SudarskyWald93} 
(described in next section) of a new and much more effective approach
(allong lines summarised in the appendix) that was initiated by
Wald in the more serene ``post classical'' era.

A similarly overhasty announcement  of my own during the hurry of the
``classical phase'' was the claim~\cite{Carter73} to have obtained an
electromagnetic generalisation of Hawking's Lichneroicz type of
staticity theorem using just the same litigious assumption (which fails
anyway for the rotating case) of the absence of an ergosphere, i.e.
strict positivity, $V>0$, of the effective gravitational potential
defined as the norm, $V=-k^\mu k_\mu$, of the stationarity generator.
What was shown later by a more careful analysis~\cite{Carter87} was that
(after correction of a sign error in the original
version~\cite{Carter73}) an even stronger and more highly litigious
inequality was in fact required -- until it was made finally redundant
by the more effective treatment recently developed by Wald and his
associates~\cite{SudarskyWald91,ChruscielWald93,SudarskyWald93} on the
lines summarised in the appendix.  (In dealing with the related
``circularity'' theorem, on which the treatment~\cite{Carter71,Carter73}
of the stationary case depends, I was more fortunate: my electromagnetic
generalisation~\cite{Carter69} of Papapetrou's pure vacuum
prototype~\cite{Papa66} has stood the test of time).

Among the other noteworthy overstatements from the hastily progressing
``classical'' period, a particularly relevant example is Wald's own
premature claim~\cite{Wald71} (based on what turns out to have been an
essentially circular argument) to have gone beyond my 1971 ``no hair''
theorem~\cite{Carter71} to get a more powerful uniqueness theorem
of the kind that was not genuinely obtained until Robinson's 1975
generalisation from infinitesimal to finite differences of the divergence
identity I had used. In achieving this ultimate {\it tour de force}
Robinson~\cite{Robinson75} effectively strengthenned the ``no hair'' theorem
to a complete uniqueness theorem, thereby definitively excluding the --
until then conceivable -- existence of a presumably unstable non-Kerr branch
of topologically spherical axisymmetric causally well behaved black hole
solutions. 

Having already succeeded in generalising my original infinitesimal
divergence identity~\cite{Carter71} from the pure vacuum to the electrovac
case~\cite{Robinson74}, Robinson went on to try to find an analogous
generalision of his more powerful finite difference divergence
identity~\cite{Robinson75} from the pure vacuum to the electrovac case.
However this turned out to be too difficult, even for him, at least by the
unsystematic, trial and error, search strategy that he and I had 
been using until then. As I guessed in a subsequent review~\cite{Carter79},
there was ``a deep but essentially simple reason why the identities found so
far should exist'' and ``the generalisation required to tie up the problem
completely will not be constructed until {\it after} the discovery of such
an expanation, which would presumably show one how to construct the required
identities {\it directly}''. It was only at a later stage, in the ``post
classical'' period that, as Heusler~\cite{Heusler96} put it `` this
prediction was shown to be true'' when, on the basis of a deeper
understanding, such direct construction methods were indeed obtained by
Mazur~\cite{Mazur82,Mazur84} and, independently, using a different
(less specialised) approach, by Bunting~\cite{Bunting83,Carter85}.

\section{The post-classical phase (since  1975).}
\label{Section4}

Robinson's 1975 discovery of the finite-difference divergence
identity~\cite{Robinson75} marked the end of what I call the ``classical
phase'', whose focal event had been the 1972 Les Houches summer
school~\cite{Hawking73,Carter73,Bardeen73} at which all the various aspects of
black hole theory were assembled and treated together for the first and
probably the last time. Since 1975 the subject has split into mutually
non-interacting branches. On one hand there has been the new subject of
quantum black hole theory: the discovery of the Hawking
effect~\cite{Hawking75} aroused interest in the possible occurrence in the
early universe of microscopic black holes for which such effects might be
important, and this in turn lead to an interest into the conceivable effects
(e.g. as potential contributors of black hole ``hair'') of various kinds of
exotic (e.g. Yang Mill or dilatonic) fields that might have been relevant
then. On the other hand, eschewing such rather wild speculations in favor of
what more obviously exists in the real world, astrophysicts have been mainly
interested in macroscopic black holes (of stellar mass and upwards) for which
the only relevant long range interaction fields are still believed to be just
gravitation and electromagnetism, but for which local mechanisms such as
accreting plasma can produce spectacular effects that are thought to be 
responsible for many observed phenomena ranging in scale and distance from 
quasars down to galactic  X-ray sources such as Cygnus X-1.

After the development of these disconnected branches, work on the pure
vacuum black hole equilibrium problem and its electrovac generalisation
proceeded rather slowly. Starting with Bekenstein~\cite{Bekenstein96},
most quantum black hole theorists were more concerned about
generalising the problem to hypothetical fields of various new (e.g.
dilatonic types), whereas most astrophysical black hole theorists were
concerned just with accreting matter that could be treated as a small
perturbation on a pure vacuum background, whose equilibrium states they
supposed to have been definitely established to consist just of the
relevant ($a\leq M$) subfamily. Only a handful of mathematically
oriented theorists remained acutely aware that the definitive
establishment of this naive supposition was still not complete. Another
reason why progress in the theory of vacuum equilibrium states slowed
down in the ``post classical phase'' was that the problems that had
been solved in the ``classical phase'' had of course tended to be those
that were easiest.

In so far as the equilibrium problem is concerned,
the most salient developments in the earlier ``post classical'' years were
the completion referred to above by Mazur~\cite{Mazur82,Mazur84} and
Bunting~\cite{Bunting83,Carter85} of my work~\cite{Carter71,Carter73} and
Robinson's~\cite{Robinson74,Robinson75} on the axisymmetric case, and the
completion and streamlining~\cite{Robinson77,Simon85,BuntingMassood87} by
Robinson, Simon, Bunting and Massood-ul-alam of the 
work~\cite{Israel67,Israel68,MRS73,MRS74} initiated by Israel
on the strictly static case. 

Unlike the work just cited, which built upwards from the (not always
entirely reliable) basis established in the classical 
period~\cite{Carter71,Hawking72,Carter73,HawkingEllis73,MRS73,MRS74}, a more
recent resurgence of  
activity~\cite{SudarskyWald91,SudarskyWald93,ChruscielWald93,ChruscielWald94,Chrusciel94,Chrusciel96} --
initiated by Wald and continued most recently by Chrusciel -- has been more
concerned with treating the shaky elements in the foundations of that
underlying basis itself. This work has successfully closed an outstanding
loophole in the previous line of argument by developing a new and more
powerful kind of staticity theorem~\cite{SudarskyWald91,SudarskyWald93} for
``non-rotating'' black holes: instead of the litigious assumption of a lower
bound on $V$ (which was needed in the now obsolete Hawking-Lichnerowicz
approach) the new theorem depends on the justifiable~\cite{ChruscielWald94}
requirement that there exists a slicing by a maximal (spacelike)
hypersurface. It has thereby been possible to
provide~\cite{ChruscielWald93,Chrusciel94,Chrusciel96} a much more more
satisfactorily complete demonstration of what had been rather
overconfidently asserted by Hawking and Ellis~\cite{HawkingEllis73}, namely
that  subject to the assumptions of analyticity and of connectedness and
non-degeneracy ($c\neq 0$) of the horizon, the black hole equilibrium state
has to be axisymmetric or static.

In so far as most of the other essential steps referred to above are
concerned, introductory presentations of the key technical details are
already available elsewhere in  surveys such as my 1987
review~\cite{Carter87} and the very extensive and up to date treatise
that has recently been provided by Heusler~\cite{Heusler96}. However
these surveys do not include any description of the technicalities of
the new improved variety of staticity
theorem~\cite{SudarskyWald91,SudarskyWald93}, whose original
presentation~\cite{SudarskyWald91,SudarskyWald93} was somewhat obscured
by extraneous complications introduced in the (so far unfulfilled) hope
of generalising the result to include Yang Mills fields. As an appendix
to the present survey, I have therefore provided a brief but self
contained account of the way this new kind of staticity theorem  is
obtained in the simplest case -- namely that of a pure (Einstein)
vacuum.

\section{What remains for the future?}

As Chrusciel has emphasised~\cite{Chrusciel96}, although many of the
(declared and hidden) assumptions involved in the work during the
``classical phase'' have been disposed of, the more recent work on the black
hole equilibrium problem is still subject to several important technical
restrictions whose treatment remains as a challenge for the future.

One whose treatment should I think be given priority at this stage is the
assumption of {\it analyticity} that has been invoked in all the work on the
indispensible  ``strong rigidity'' theorem that is needed to establish
axisymmetry. It is to be remarked that if, as in my early
work~\cite{Carter71}, axisymmetry is simply postulated at the outset, then
analyticity will be demonstrable as an automatic consequence of the
ellipticity of the differential system that is obtained as a result of the
``circularity'' property that is established by the generalised Papapetrou
theorem~\cite{Carter69,Carter71}. What I would guess is that it should be
possible (and probably not more difficult than the other steps that have
already been acheived) to prove the necessity of analyticity for a vacuum
equilibrium state {it without} assuming axisymmetry.

A more delicate question that remains to be settled is the possibility
of equilibrium involving {\it several disconnected black holes}.  As
far as the pure vacuum problem is concerned, my conjecture is that such
multi black hole solutions do not exist, but they have so far been
rigorously excluded only in the strictly static
case~\cite{BuntingMassood87}. The axisymmetric case has recently been
studied in some detail~\cite{Weinstein90,Weinstein94} by Weinstein (who
denotes the horizon scale parameter $c=\kappa{\cal A}/4\pi$ by the
letter $\mu$) but a definitive conclusion has not yet emerged. In the
electrovac case the situation is certainly more complicated, since it
is known~\cite{HartleHawking72} that there are counterexamples in which
gravitational attraction is balanced by electrostatic repulsion. It is
however to be noticed that the only counterexamples discovered so far,
namely those of the Papapetrou-Majumdar family~\cite{Papa45,Majumdar47}
have horizons that are degenerate (in the sense of having a vanishing
decay constant $\kappa$). It seems reasonable to conjecture that even
in the electrovac case there are no non-degenerate multi black hole
equilibrium states.

This last point leads on to a third major problem that still needs to be
dealt with, namely the general treatment of the {\it degenerate} ($\kappa=0$)
case. The maximally rotating ($J^2=M^4$) Kerr solution is still the only known
pure vacuum example, and I am still inclined to conjecture that it is unique,
but the problem of proving this remains entirely unsolved. As far as the
electrovac problem is concerned, the only known examples are those of the
Kerr-Newman~\cite{Newman65} and Papapetrou-Majumdar~\cite{Papa45,Majumdar47}
families. Recent progress by Heusler~\cite{Heusler97} has confirmed that the
latter (whose equilibrium saturates a Bogomolny type mass limit~\cite{GHHP83})
are the only strictly static examples, but for the rotating degenerate case
the problem remains wide open.

I wish to conclude by drawing attention to another deeper problem that,
unlike the three referred to above, has been largely overlooked even by
the experts in the field, but that seems to me just as interesting from
a purely mathematical point of view, even if its physical relevance is
less evident. This fourth problem, is that of solving the black hole equilibrium state problem
without invoking the {\it causality axiom} on which nearly all the work
described above depends (e.g. for obtaining the required positivity in the
successive divergence 
identities~\cite{Carter71,Robinson74,Robinson75,Bunting83,Mazur82,Mazur84,Carter85} 
used in the axisymmetric case). As remarked in the introduction, all non
static Kerr solutions contain closed timelike lines, though in the black hole
subfamily with $J^2\leq M^4$ they are entirely confined inside the
horizon~\cite{Carter68,Carter78}. Unlike analyticity, whose failure in shock
type phenomena is physically familiar in many contexts, causality -- meaning
the absence of closed timelike lines -- is a requirement that most physicists
would be prepared to take for granted as an indispensible requirement for
realism in any classical field model. However the example of sphalerons
suggests that despite their unacceptability at a classical level, the
mathematical existence of stationary black hole states with closed timelike
lines outside the horizon might have physically relevant implications in
quantum theory. The discovery of such exotic configurations would be a
surprise to most of us, but would not contradict any theorem obtained so far.
All that can be confidently asserted at this stage is that such configurations
could not be static but would have to be of rotating type.


\section*{Appendix: the new staticity theorem.}

In view of the importance of the new kind of staticity theorem
developed~\cite{SudarskyWald91,SudarskyWald93} by Sudarsky and Wald
(superceding the Lichnerowicz kind, whose adaptation to the black hole
context was inadequate  in the pure vacuum case dealt with by
Hawking~\cite{HawkingEllis73}, and even less satisfactory in the
electromagnetic case dealt by myself~\cite{Carter73,Carter87}), this
appendix presents a brief but self contained summary of the essential
ideas.  The Sudarsky Wald approach works
perfectly well for the electrovac case (though it does seem to have
trouble with Yang Mills fields) but in order to display the key points
in the simplest possible form the description below is limited to the
case of a pure (Einstein) vacuum.

It is first to be recalled that the vacuum black hole equilibrium
configurations under consideration belong to the more general category
of equilibrium configurations, including those of isolated states of
self gravitating bodies such as neutron star models, that are
invariant under the action of a time translation
group whose generator $k^\mu$ is timelike at least at
large sufficiently large distance in the asymptotically flat outer
region -- where it can be taken to be normalised so that its magnitude
tends to unity at large distance.  Any such configuration will of
course have a well defined assymptotic mass, $M$ say, given by a
formula of the standard Komar form
 \be  M={1\over 4\pi}\int_\infty k^{\mu;\nu}\, d\!{\cal S}_{\mu\nu} \,
 , \eqn{Komarmass}\fe
 (using a semi-colon to indicate covariant differentiation) where the
integral is taken over a surrounding topologically spherical spacelike
2-surface ${\cal S}_\infty$ whose choice is arbitrarily adjustable
without affecting the result provided it is taken sufficiently far out
to be entirely in the vacuum region where the source free Einstein
equations are satisfied. In the pure vacuum black hole case with which
we are concerned here, the analogous integral defining the black hole
mass contribution
 \be M_{\cal H}={1\over 4\pi}\int_{\cal H} k^{\mu;\nu}\,
 d\!{\cal S}_{\mu\nu} \, ,\eqn{holemass}\fe 
in terms of any spacelike 2-surface ${\cal S}_{\cal H}$ on the horizon
will give the same result:  the vanishing of the Riemann tensor
$R_{\mu\nu}$ in conjunction with the Killing equation \be
k_{(\mu;\nu)}=0 \eqn{Killing}\fe
 (using round brackets for ansymmetrisation) ensures that divergence
condition $k^{\mu;\nu}{_{;\nu}} =0$ is satisfied all the way in to the
horizon, with the implication that that $M_{\cal  H}=M$.

According to Hawking's ``strong rigidity theorem''
~\cite{HawkingEllis73} (which depends on the not yet satisfactorily
justified analyticity postulate~\cite{Chrusciel94} dicussed above) 
the null tangent vector of the horizon will be normalisable in such a way as
to coincide with a ``corotating'' Killing vector field $\ell^\mu$ given by a
formula of the standard form
 \be  \ell^\mu=k^\mu+\Omega^{\cal H} h^\mu \, ,\eqn{rigidity}\fe
where $\Omega^{\cal H}$ is a {\it uniform} (``rigid'') angular velocity
and $h^\mu$ is indeterminate if $\Omega^{\cal H}=0$ (i.e. in the
``non-rotating'' case) and otherwise is a well defined (\it
axisymmetry} Killing vector -- with circular trajectories such that the
correspondingly normalised angle parameter has period
$2\pi$.  (If, instead of assuming analyticity, one assumes the existence
of the axisymmetry generated by $h^\mu$ then the
formula (\ref{rigidity}) is very easily derivable by my ``weak
rigidity'' theorm~\cite{Carter72}). The  scale constant $c$ of the
horizon is defined by an expression of the same form as that for the
mass but with the original (asymptotically timelike) Killing vector
$k^\mu$ replaced by the new the Killing vector combination $\ell^\mu$
that is null on the horizon, i.e.  it is specified by
  \be  c={1\over 4\pi}\int_{\cal H} \ell^{\mu;\nu}\, d\!{\cal S}_{\mu\nu}
  \, ,\eqn{holescale}\fe 
In terms of the acceleration parameter
${\kappa}$ given by $2\kappa^2=\ell^{\mu;\nu} \ell_{\nu;\mu}$ which
(like $\Omega^{\cal H}$) must be {\it uniform} over the horizon by the
``zeroth'' law~\cite{BCH73}, the scale constant works out locally
~\cite{Carter73} as
 \be c={\kappa{\cal A}\over 4\pi} \, , \eqn{scalevalue}\fe 
where ${\cal A}$ is the horizon area. By the ``rigidity''
formula (\ref{rigidity}) (whether obtained from the ``weak
theorem''~\cite{Carter72} assuming axisymmetry, or from the ``strong
theorem''~\cite{HawkingEllis73} assuming analyticity) it immediately follows 
that the scale parameter will be expressible in terms of globally defined 
quantities via the Smarr type relation
 \be c=M_{\cal H}-2\Omega^{\cal H} J_{\cal H}\, , \eqn{locSmarr}\fe
where $M_{\cal H}$ is the black hole mass contribution as defined above
and $J_{\cal H}$ is the corresponding black hole angular momentum
contribution,
 \be J_{\cal H}=-{1\over 16\pi} \oint_{\cal H} h^{\mu;\nu} \, d\!{\cal
 S}_{\mu\nu} \, , \eqn{holeangm} \fe 
which will be the same as the {\it total} angular momentum
 \be J =-{1\over 16\pi} \oint_\infty  h^{\mu;\nu} \, d\!{\cal
 S}_{\mu\nu} \, , \eqn{totangm} \fe 
in the pure vacuum case considered here, i.e. we shall have  $J_{\cal H}=J$.

The new idea in the relatively recent work of Wald and
Sudarsky~\cite{SudarskyWald91,SudarskyWald93} is to compare the long
well known formula (\ref{locSmarr}) that has just been recapitulated,
which in the pure vacuum case under consideration here is evidently
equivalent to the simple global mass formula
 \be M=c+2\Omega^{\cal H} J \, ,\eqn{totSmarr}\fe
with a mass formula of the  Arnowitt Deser Misner
kind, which involves integration over a spacelike 3-surface $\Sigma $ say.
The trick used by Wald and Sudarsky was to choose the
hypersurface $\Sigma$ to be {\it maximal} -- a restriction that has been
confirmed to be imposable without loss of generality by Wald and
Chrusciel~\cite{ChruscielWald93}. This means that its second fundamental form,
as expressed (using Latin letters for internal coordinate indices on
the hypersurface) by $K_{ij}$ should be trace free, i.e. $K_i^{\ i}=0$ 
(using the induced 3-metric. i.e. the first fundamental form, for index
raising). In the vacuum case this reduces the A.D.M. formula to 
the simple form
 \be M={1\over 4\pi}\int_{\Sigma}K^{ij}K_{ij}\lambda\, d\Sigma +{1\over
 4\pi} \oint_{\cal H} \lambda_{,i} \, d\! S^i \eqn{Waldmass}\fe where
the surface field $\lambda$ is given in terms of the unit normal
$n^\mu$ to $\Sigma$ by $\lambda=-k^\mu n_\mu$ (which will be
positive).  Since the axisymmetry generator $h^\mu$ will be tangential
to a section $\Sigma$ that is maximal, one will have $h^\mu n_\mu=0$,
so that the ``rigidity'' formula (\ref{rigidity}) will simply give
$\lambda= -\ell^\mu n_\mu$ on the horizon. The boundary 2-surface
contribution from the horizon can thus be evaluated as
 \be {1\over 4\pi} \oint_{\cal H} \lambda_{,i}\, d\! S^i= c
\, , \eqn{holecon} \fe where $c$ is the scale parameter as defined by
(\ref{holescale}).

Identifying the output of this A.D.M. type mass formula (\ref{holecon})  
with that of our older Smarr type formula (\ref{totSmarr}), one
obtains a relationship expressible (in the pure vacuum case
under consideration here) by
 \be M-c=2 \Omega^{\cal H} J= {1\over 4\pi}\int_\Sigma K^{ij} K_{ij}
 \lambda \, d\Sigma\, . \eqn{Waldid} \fe
The manifest non-negativity of the integrand on the right hand side
of the Wald Sudarsky identity (\ref{Waldid}) evidently entails that
in the non rotating case the second fundamental form must vanish, i.e.
 \be  \Omega^{\cal H} =0\hskip 1cm \Rightarrow\hskip 1cm  K_{ij}=0
\, .\eqn{intflat} \fe
Having thus established the extrinsic flatness of the maximal hypersurface
for the case of a stationary black hole with non-rotating horizon, one can
straightforwardly proceed to show that as a consequence the vector $t^\mu$
defined by 
 \be t^\mu=\lambda n^\mu  \eqn{normdef}\fe
will automaticall satisfy a Killing equation $t_{(\mu;\nu)}=0$ of the same
form as the one (\ref{Killing}) satisfied by the original time translation
generator,  with which it will therefore be identifiable, i.e. one obtains  
 \be t^\mu=k^\mu\, .\eqn{idi}\fe
Since $t^\mu$ is hypersurface orthogonal by its
construction (\ref{normdef}), the desired staticity theorem is
thereby established: it has been shown that the vanishing of
the black hole angular velocity  $\Omega^{\cal H}$ is sufficient
{\it by itself} (without the need to postulate a questionable
lower limit on the magnitude $V=-k^\mu k_\mu$ as in the older treatment) to
ensure hypersurface orthogonality of the time translation symmetry
generator $k^\mu$.



\begin{thebibliography}{99}


\bibitem{Carter67} B. Carter, ``Stationary Axisymmetric Systems in General
Relativity" (Ph.D. Thesis, DAMTP, Cambridge, 1977).

\bibitem{Carter68} ``Global Structure of the Kerr Family of
Gravitational Fields", B. Carter, {\it Phys. Rev.} {\bf 174}, 
1559-71 (1968).

\bibitem{Kerr63} R.P. Kerr, ``Gravitational field of a spinning mass as an
example of algebraically special metrics'',
{\it Phys. Rev. Lett.} {\bf 11}, 237-38 (1963).

\bibitem{Israel67} W. Israel, ``Event horizons in static vacuum 
spacetimes'', {\it Phys.Rev} {\bf 164}, 1776-79 (1967).

\bibitem{Carter71} B. Carter, ``An Axisymmetric Black Hole has only Two
Degrees of Freedom", B. Carter, {\it Phys. Rev. Letters} {\bf 26}, 
331-33 (1971).

\bibitem{Hawking72} S.W. Hawking ``Black holes in General Relativity'',
{\it Commun. Math. Phys.} {\bf 25}, 152-56 (1972).

\bibitem{Hawking73} S.W. Hawking, ``The event horizon'', 
in {it Black Holes}, (proc. 1972 Les Houches Summer School), 
ed. B. \& C. DeWitt, 1-55 (Gordon and Breach, New York, 1973).
 
\bibitem{Carter73} B. Carter, ``Black Hole Equilibrium States: II
General Theory of Stationary Black Hole States", in {\it Black Holes}
(proc. 1972 Les Houches Summer School), ed. B. \& C. DeWitt, 
125-210 (Gordon and Breach, New York, 1973).

\bibitem{HawkingEllis73} S.W. Hawking, G.F.R. Ellis, {\it The Large
Scale Structure of Space Time}, (Cambridge U.P., 1973).

\bibitem{Robinson74} D.C. Robinson,  ``Classification of black holes with 
electromagnetic fields'', {\it Phys, Rev.} {\bf D10}, 
458-60 (1974)

\bibitem{Robinson75} D.C. Robinson, ``Uniqueness of the Kerr black hole'', 
{\it Phys, Rev. Lett.}{\bf 34}, 905-06 (1975)

\bibitem{Robinson77} D.C. Robinson, ``A simple proof of the generalisation
of Israel's theorem'', {\it Gen. Rel. Grav.} {\bf 8},  
695-98 (1977).

\bibitem{Bunting83} G. Bunting, ``Proof of the Uniqueness Conjecture 
for Black Holes,''(Ph.D. Thesis, University of New England, 
Armadale N.S.W., 1983).

\bibitem{Carter85} B. Carter, ``The Bunting Identity and Mazur Identity 
for non-linear Elliptic Systems including the Black Hole Equilibrium 
Problem", {\it Commun. Math. Phys.} {\bf 99}, 563-91 (1985).

\bibitem{Mazur82} P.O. Mazur, ``Proof of uniqueness of the Kerr-Newman
black hole solution'', {\it J. Phys.} {\bf A15}, 3173-80 (1982).

\bibitem{Mazur84} P.O. Mazur, ``Black hole uniqueness from a hidden symmetry 
of Einstein's gravity'', {\it Gen. Rel Grav.} {\bf 16}, 211-15
(1984).

\bibitem{Carter87} B. Carter ``Mathematical foundations of the theory of
relativistic stellar and black hole configurations'', in {\it Gravitation
in Astrophysics}  (NATO ASI {\bf B156}, Carg\`ese, 1986), ed. B. Carter, 
J.B.~Hartle, 63-122 (Plenum Press, New York, 1987).


\bibitem{SudarskyWald91} D. Sudarsky, R.M. Wald, ``Extrema of mass, 
stationarity, and staticity, and solutions to the Einstein-Yang-Mills 
equations'', {\it Phys. Rev.} {\bf D46}, 1453-74 (1991)

\bibitem{ChruscielWald93} P.T Chrusciel, R.M. Wald, ``Maximal 
hypersurfaces in stationary asymptotically flat spacetimes'',
{\it Comm. Math. Phys.} {\bf 163}, 561-604 (1964).
[gr-qc/9304009]

\bibitem{SudarskyWald93} D. Sudarsky, R.M. Wald, ``Mass formulas for 
stationary Einstein-Yang-Mills black holes and a simple proof of two 
staticity theorems'', {\it Phys. Rev.} {\bf D47}, 5209-13 (1993). 
[gr-qc/9305023]

\bibitem{ChruscielWald94} P.T. Chrusciel, R.M. Wald, ``On the topology
of stationary black holes'', {\it Class. Quantum Grav.} {\bf 11},
L147-52 (1994).
[gr-qc/9410004]

\bibitem{Thorne94} K.S. Thorne, ``Black holes and time warps''
(Norton, New York, 1994).

\bibitem{Israel87}  W. Israel, ``Dark stars: the evolution of an idea'',
 in {\it 300 years of gravitation}, ed. S.W. Hawking, W. Israel, 
199-276 (1987).

\bibitem{Hawking75} S.W. Hawking, ``Particle creation by black holes'',
{\it Comm. Math. Phys} {\bf 43}, 199-220 (1975).

\bibitem{Chrusciel94} P.T. Chrusciel, ``No-hair theorems: folkelore, 
conjectures, results'', in {\it Differential Geometry and Mathematical 
Physics} {\bf 170}, ed J.Beem, K.L. Dugal, 23-49 (American 
Math. Soc., Providence, 1994) 
[gr-qc/9402032]

\bibitem{Heusler96} M. Heusler, {\it Black hole uniqueness theorems}
(Cambridge U.P., 1996) 

\bibitem{Carter78} B.Carter, ``Domains of stationary communications'',
{\it Gen. Rel. Grav.} {\bf 11}, 437-50 (1978).

\bibitem{Schwarz16} ``Uber das Gravitationsfeld eines Massenpunctes
nach der Einsteinschen Theorie'' {\it Sitzber. Deut. Akad. Wiss. Berlin Kl.
Math-Phys. Tech.}, 189-96 (1916)

\bibitem{Eisenstaedt82} J. Eisenstaedt,``Histoire et singularit\'es de la 
solution de Schwarzschild (1915-1923)'', {\it Arch. Hist. Exact Sci} 
{\bf 27}, 157-228 (1982).

\bibitem{Birkhoff23} G.D. Birkhoff, {\it Relativity and Modern Physics} 
(Harvard U.P., 1923).

\bibitem{Oppy39} J.R. Oppenheimer, H. Snyder ``On continued gravitational 
contraction'', {\it Phys. Rev.} {\bf 56}, 455-59 (1939).

\bibitem{Painleve21}  P. Painlev\'e, ``La m\'ecanique classique et la
theorie de la relativit\'e'', {\it C.R. Acad. Sci. (Paris)} {\bf 173},
677-80 (1921).

\bibitem{Gullstrand22} A. Gullstrand, ``Allegemeinne Losung des statischen
Einkorper-problems in der Einsteinschen Gravitations theorie'',
{\it Archiv. Mat. Astron. Fys.} {\bf 16(8)}, 1-15 (1922).

\bibitem{Eddington24} A.S. Eddington, ``A comparison of  Whitehead's and 
Einstein's formulas'', {\it Nature} {\bf 113}, 192 (1924).

\bibitem{Lemaitre30}  G. Lemaitre, ``L'univers en expansion'',
{\it Ann. Soc. Sci. Bruxelles} {\bf I A53}, 51-85 (1933). 

\bibitem{Synge50} J.L. Synge ``The gravitational field of a particle'',
{\it Proc. R. Irish Acad.} {\bf A53}, 83-114 (1950).

\bibitem{Finkelstein58} D. Finkelstein ``Past-future asymmetry of a point 
particle'', {\it Phys. Rev.} {\bf 110} 965-67 (1958).

\bibitem{Frondsal59} C. Frondsal ``Completion and embedding of the
Schwarzschild solution'', {\it Phys. Rev.} {\bf 116}, 778-81 
(1959).

\bibitem{Kruzkal60} M.D. Kruskal, ``Minimal extension of the Schwarzschild
metric'', {\it Phys. Rev.} {\bf 119}, 1743-45 (1960).

\bibitem{Szekeres60} G. Szekeres, ``On the singularities of a Riemannian 
manifold'', {\it Publ. Math. Debrecen} {\bf 7}, 285-301 (1960)

\bibitem{ZeldovichNovikov67} Ya. B. Zel'dovich, I.D. Novikov,
{\it Relativistic Astrophysics} (Izdatel'stvo ``Nauka'', Moscow 1967); 
English version ed. K.S Thorne, W.D. Arnett (University of 
Chicago Press, 1971).

\bibitem{HTWW65} B.K. Harrison, K.S. Thorne, M. Wakano, J.A. Wheeler,
{it Gravitation theory and gravitational collapse} (University of
Chicago Press, 1965).

\bibitem{Penrose65} R. Penrose, ``Gravitational collapse and spacetime 
singularities'', {\it Phys. Rev. Letters} {\bf 14}, 57-59 (1965).

\bibitem{Vishu68} C.V. Vishveshwara, ``Generalisation of the `Schwarzschild 
surface' to arbitrary static and stationary spacetimes'', 
{\it J. Math. Phys} {\bf 9}, 1319-22 (1968).

\bibitem{Israel68} W. Israel, ``Event horizons in static electrovac 
spacetimes'', {\it Commun. Math. Phys.,} {\bf 8}, 245-60 (1968)

\bibitem{Boyer65} R.H. Boyer, T.G. Price, ``An interpretation of the Kerr 
metric in General Relativity'', {\it Proc. Camb.Phil.Soc.} 
{\bf 61}, 531-34 (1965).

\bibitem{Boyer67} R.H. Boyer, R.W. Lindquist, ``Maximal analytic extension
of the Kerr metric'', {\it J. Math. Phys.} {\bf 8}, 265-81 (1967).

\bibitem{Boyer69} R.H. Boyer, ``Geodesic orbits and bifurcate Killing
horizons'', {\it Proc. Roy. Soc. Lond} {\bf A311}, 245-52 (1969). 

\bibitem{Carter66} B. Carter, ``Complete analytic extension of the
symmetry axis of Kerr's solution of Einstein's Equations", {\it Phys.
Rev.} {\bf 141}, 1242-47 (1966)

\bibitem{Bardeen73} J. Bardeen, ``Rapidly rotating stars, disks, and black 
holes'', in {it Black Holes}, (proc. 1972 Les Houches Summer School), 
ed. B. \& C. DeWitt, 241-89 (Gordon and Breach, New York, 1973).

\bibitem{HartleHawking72} J.B. Hartle, S.W. Hawking, ``Solutions of the 
Einstein-Maxwell equations with many black holes'', 
{\it Commun. Math. Phys.} {\bf 26}, 87-101 (1972)

\bibitem{Newman65} E.T. Newman, E. Couch, E. Chinnapared, K. Exton,
AQ. Prakash, R. Torrence, ``Metric of rotating charged mass'',
{\it J. Math. Phys.} {\bf 6} 918-19 (1965).

\bibitem{Papa45} A. Papapetrou, ``A static solution of the gravitational
field for arbitrary charge distribution'', {Proc. R. Irish Acad.} 
{\bf A51}, 191-204 (1945)

\bibitem{Majumdar47} S.D. Majumdar, ``A class of exact solutions of 
Einstein's field equations'', {\it Phys. Rev.} {\bf 72}, 
930-98 (1947).

\bibitem{Penrose69} R. Penrose, ``Gravitational collapse: the role of
general relativity'', {\it Riv. Nuovo Cimento I} {\bf 1}, 
252-76 (1969).

\bibitem{Eardley87} D. Eardley, ``Naked singularities in spherical 
gravitational collapse'', in {\it Gravitation in Astrophysics}  
(NATO ASI {\bf B156}, Carg\`ese, 1986), ed. B. Carter, 
J.B.~Hartle, 229-35 (Plenum Press, New York, 1987).

\bibitem{Vishu70} C.V. Vishveshwara, ``Stability of the Schwarzschild 
metric'', {\it Phys. Rev.} {\bf D1} 2870-79 (1970).

\bibitem{Price72} R.H. Price. `` Non spherical perturbations of 
relativistic collapse, I : scalar and gravitational perturbations; 
II: integer spin zero rest mass fields '', {\it Phys. Rev.} 
{\bf D5} 2419-54 (1972).

\bibitem{Teukolsky72} S.A. Teukolsky, ``Rotating black holes:
separable wave equations for gravitational and electromagnetic
perturbations '',
{\it Phys. Rev. Lett.} {\bf 29}, 1114-18 (1972).

\bibitem{Teukolsky73}  S.A. Teukolsky, ``Perturbations of a rotating black 
hole, I: Fundamental equations for gravitational and electromagnetic 
perturbations '', {\it Astroph. J.} {\bf 185}, 635-47 (1973).
		  
\bibitem{PressTeukolsky73} W.H. Press, S.A. Teukolsky, ``Perturbations 
of a rotating black  hole, II: Dynamical stability of the Kerr metric '',       
{\it Astroph. J.} {\bf 185}, 649-73 (1973).

\bibitem{KayWald87} B.S. Kay, R.M. Wald, ``Linear stability of Schwarzschild 
under perturbations which are nonvanishing on the bifurcation 2-sphere''
{\it Class. Quantum Grav.} {\bf 4}, 893-98 (1987).

\bibitem{Whiting89} B. Whiting, ``Mode stability of the Kerr blackhole'',
{\it J. Math. Phys.} {\bf 30}, 1301-05 (1989).

\bibitem{Christodoulou70} D. Christodoulou, ``Reversible and irreversible
transformations in black hole physics'', {\it Phys. Rev. Lett.}
{\bf 25}, 1596-97 (1970)

\bibitem{ChristodoulouRuffini71} D. Christodoulou, R. Ruffini, ``Reversible
transformations of a charged black hole'',
{\it Phys. Rev.} {\bf D4}, 3552-55 (1971).

\bibitem{Carter72} B. Carter, ``Rigidity of a black hole'' {\it Nature}
{\bf 238 }{Physical Science}, 71-72 (1972).

\bibitem{HawkingHartle72} S.W. Hawking, J.B. Hartle, ``Energy and momentum 
flow into a black hole'',
{\it Commun. Math. Phys.} {\bf 27}, 283-90 (1972) 

\bibitem{Hartle73} J.B. Hartle, ``Tidal friction in slowly rotating black 
holes'', {\it Phys. Rev.} {\bf D8}, 1010-24 (1973).

\bibitem{BCH73} J. Bardeeen, B. Carter, S.W. Hawking, ``The four laws 
of black hole mechanics'', {\it Comm. Math. Phys.} {\bf 31},
161-70 (1973).

\bibitem{Znajek78}  R.L. Znajek, ``Charged current loops around Kerr
holes'', {\it Mon. Not. R. Astr. Soc.} {\bf 182}, 639-46 (1978).

\bibitem{Damour78} T. Damour, ``Black hole eddy currents'', 
{\it Phys. Rev.} {\bf D18}, 3598-604 (1978).

\bibitem{Carter79} B. Carter, ``The general theory of the mechanical,
electromagnetic, and thermodynamical properties of black holes'', in {\it
General Relativty, an Einstein centenary survey}, ed. S.W. Hawking, W.
Israel, 294-369 (Cambridge U.P., 1979).

\bibitem{Bekenstein73} J.D. Bekenstein, ``Extraction of energy and charge
from a black hole'', {\it Phys. Rev.} {\bf D7}, 949-53 (1973).

\bibitem{Ernst68} F.J. Ernst,  ``New formulation of the axially symmetric
gravitational field problem'' {\it Phys. Rev.} {\bf 167}, 1175-78
 and {\bf 168}, 1415-17 (1968).

\bibitem{MRS73} H. Muller zum Hagen, D.C. Robinson, H.J. Seifert,
``Black Holes in static vacuum spacetimes'', {\it Gen. Rel. Grav.} 
{\bf 4}, 53-78 (1973).

\bibitem{MRS74}H. Muller zum Hagen, D.C. Robinson, H.J. Seifert,
``Black Holes in static electrovacspacetimes'', {\it Gen. Rel. Grav.} 
{\bf 5}, 61-72 (1974).

\bibitem{Lichnerowicz55} A. Lichnerowicz, {\it Th\'eories relativistes
de la gravitation et de l'electro-magn\'etisme} (Masson, Paris, 1955).

\bibitem{Wald71} R.M. Wald, ``Final states of gravitational collapse'', 
{\it Phys. Rev. Lett.} {\bf 26}, 1653-55 (1971)

\bibitem{Bekenstein96} J.D. Bekenstein ``Black hole hair: twenty five
years after'', preprint (Hebrew University, Jerusalem, 1996)
[gr-qc/9605059]

\bibitem{Carter69} B. Carter, ``Killing horizons and orthogonally
transitive groups in space-time'', {\it J. Math. Phys.} {\bf 10},
1559-71 (1969).

\bibitem{Papa66} A. Papapetrou, ``Champs gravitationnels
stationnaires \`a symmetrie axiale'', {\it Ann. Inst. H. Poincar\'e}
{\bf 4}, 83-85 (1966).

\bibitem{Simon85} W. Simon, ``A simple proof of the generalised Israel
theorem'', {\it Gen. Rel. Grav.} {\bf 17}, 761-68 (1985).


\bibitem{BuntingMassood87} G.L. Bunting, A.K.M. Massood-ul-Alam,
``Nonexistence of multiple black holes in   asymptotically Euclidean
static vacuum space-times'', {\it Gen. Rel. Grav.} {\bf 19}, 
147-54 (1987).

\bibitem{Chrusciel96} P.T. Chrusciel ``Uniqueness of stationary
electrovacuum black holes revisited'', {\it Helv. Phys. Acta}
{\bf 69}, 529-52 (1996).

\bibitem{Weinstein90} G. Weinstein, ``The stationary axisymmetric two
body problem in general relativity'', {\it Comm. Pure. Appl. Math}
{\bf XLV} 1183-203 (1990).

\bibitem{Weinstein94} G. Weinstein, ``On the force between rotating
coaxial black holes'', {\it Trans. Amer. Math. Soc.} {\bf 343},
899-906 (1994).

\bibitem{Weinstein96} G. Weinstein, ``N-black hole 
stationary and axially symmetric solutions of the Einstein-Maxwell
equations'' {\it Comm. Part. Diff. Eq.}, in press (1996).
[gr-qc/9412036]

\bibitem{Heusler97} M. Heusler, ``On the uniqueness of the 
Papapetrou-Majumdar metric'', preprint (University of Zurich, 1996).
[gr-qc/9607001]

\bibitem{GHHP83} G.W. Gibbons, S.W. Hawking, G.T. Horowicz, M.J. Perry,
``Positive mass theorem for black holes'',
{\it Comm. Math. Phys.} {\bf 88}, 259-308 (1983).

\end{thebibliography}
\end{document}